\documentclass[10pt,conference]{IEEEtran}
\IEEEoverridecommandlockouts
\usepackage{cite}
\usepackage{amsmath,amssymb,amsfonts}
\usepackage{algorithmic}
\usepackage{graphicx}
\usepackage{textcomp}
\usepackage{xcolor}
\usepackage{subcaption}
\usepackage{multirow}
\usepackage{tablefootnote}
\usepackage[framemethod=tikz]{mdframed}
\usepackage{adjustbox}
\usepackage{soul}
\usepackage{listings}
\usepackage{xcolor}

\usepackage{tikz}
\usepackage{amsmath}
\usepackage{multicol}
\usepackage{caption}
\usetikzlibrary{decorations.pathreplacing} % Add this line

\usepackage{xcolor}  % For color customization

\usepackage{stfloats} 
\usepackage{tabularx}
\usepackage{booktabs}

\definecolor{personaColor}{HTML}{FFDD57}  % Light yellow
\definecolor{objectiveColor}{HTML}{57FFDD}  % Light cyan
\definecolor{contextColor}{HTML}{FF57DD}  % Light pink
\definecolor{taskDescColor}{HTML}{57DDFF}  % Light blue
\definecolor{outputInstColor}{HTML}{DD57FF}  % Light purple
\definecolor{instructionColor}{HTML}{DDFF57}  % Light lime
\definecolor{exampleColor}{HTML}{FFA500}  % Light orange

\def\BibTeX{{\rm B\kern-.05em{\sc i\kern-.025em b}\kern-.08em
    T\kern-.1667em\lower.7ex\hbox{E}\kern-.125emX}}
\begin{document}

\title{Prompting in the Wild: An Empirical Study of Prompt Evolution in Software Repositories}

\makeatletter
\newcommand{\linebreakand}{%
  \end{@IEEEauthorhalign}
  \hfill\mbox{}\par
  \mbox{}\hfill\begin{@IEEEauthorhalign}
}
\makeatother

\author{
    \IEEEauthorblockN{Mahan Tafreshipour}
    \IEEEauthorblockA{
        University of California, Irvine \\
        Irvine, USA \\
        mtafresh@uci.edu
    }
    \and
    \IEEEauthorblockN{Aaron Imani}
    \IEEEauthorblockA{
        University of California, Irvine \\
        Irvine, USA \\
        aaron.imani@uci.edu
    }
    \and
    \IEEEauthorblockN{Eric Huang}
    \IEEEauthorblockA{
        University of California, Irvine \\
        Irvine, USA \\
        jianch2@uci.edu
    }
    \linebreakand
    \IEEEauthorblockN{Eduardo Almeida}
    \IEEEauthorblockA{
        Federal University of Bahia (UFBA) \\
        Brazil \\
        esa@rise.com.br
    }
    \and
    \IEEEauthorblockN{Thomas Zimmermann}
    \IEEEauthorblockA{
        University of California, Irvine \\
        Irvine, USA \\
        tzimmer@uci.edu
    }
    \and
    \IEEEauthorblockN{Iftekhar Ahmed}
    \IEEEauthorblockA{
        University of California, Irvine \\
        Irvine, USA \\
        iftekha@uci.edu
    }
}
\maketitle

\begin{abstract}
%OLD: The adoption of Large Language Models (LLMs) is transforming software development as developers incorporate these tools into their applications and produce LLM-integrated applications. Prompts are the initial means of interaction with LLMs in such applications. Despite the widespread use of LLM-integrated applications, there is still a limited understanding of how developers manage and evolve prompts in these applications. In this study, we present the first empirical analysis of prompt evolution in LLM-integrated software development. We analyzed 1,262 prompt changes across 243 GitHub repositories to explore the patterns and frequencies of prompt changes, their relationship with code changes, documentation practices, and their impact on system behavior. Our findings reveal that developers primarily evolve prompts through additions and modifications, with most changes occurring during feature development. We identified critical challenges in prompt engineering practices: only 21.9\% of prompt changes are documented in commit messages, changes can introduce logical inconsistencies in prompts, and there is a misalignment between prompt changes and LLM responses. These insights highlight the need for specialized testing frameworks, automated validation tools, and improved documentation practices to enhance the reliability of LLM-integrated applications.

The adoption of Large Language Models (LLMs) is reshaping software development as developers integrate these LLMs into their applications. In such applications, prompts serve as the primary means of interacting with LLMs. Despite the widespread use of LLM-integrated applications, there is limited understanding of how developers manage and evolve prompts. This study presents the first empirical analysis of prompt evolution in LLM-integrated software development. We analyzed 1,262 prompt changes across 243 GitHub repositories to investigate the patterns and frequencies of prompt changes, their relationship with code changes, documentation practices, and their impact on system behavior. Our findings show that developers primarily evolve prompts through additions and modifications, with most changes occurring during feature development. We identified key challenges in prompt engineering: only 21.9\% of prompt changes are documented in commit messages, changes can introduce logical inconsistencies, and misalignment often occurs between prompt changes and LLM responses. These insights emphasize the need for specialized testing frameworks, automated validation tools, and improved documentation practices to enhance the reliability of LLM-integrated applications.

% The adoption of large language models (LLMs) is transforming software development as developers incorporate these tools into their applications, creating LLM-integrated systems. Prompts serve as the primary interface for interacting with LLMs to perform specific tasks. Despite the widespread use of such applications, there is limited understanding of how developers manage and refine prompts over time to meet evolving project needs. In this study, we present the first empirical analysis of prompt evolution within LLM-integrated software development. Analyzing 1,262 prompt changes across 243 diverse GitHub repositories, we explore the patterns and frequencies of these changes, their relationship with code changes, documentation practices, and their impact on application behavior. Our findings indicate that prompt evolution is an iterative and relatively stable process compared to code changes. However, significant gaps exist in prompt documentation, and there are inconsistent alignments between intended prompt modifications and LLM responses. These insights highlight the need for better prompt engineering tools, improved documentation standards, and robust prompt validation methods to increase the reliability of LLM-integrated applications.
\end{abstract}


\section{Introduction}
\label{sec:intro}

% what is the main topic of this para? write topic sentences for each para
With the rise of Large Language Models (LLM), software developers and practitioners have increasingly adopted these tools as software components in their applications. LLM-integrated applications, also referred to as FMwares (Foundational Model ware), are software systems that interact with LLMs via API calls and use the generated responses to perform specific tasks \cite{10.1145/3663529.3663849, Weber2024LargeLM}. These tasks encompass a broad spectrum of functionalities, ranging from the development of new features such as question-answering chatbots to the automation of existing processes, like SQL query generation, or even the construction of entire applications centered around an LLM\cite{nahar2024beyond}. A prominent example of this adoption is OpenAI's GPT applications, which are simple applications built in the ChatGPT environment and which have gained substantial traction, with over 3 million custom GPTs published in the GPTStore as of January 2024 \cite{Baek2024}.  Many of these applications extend beyond simple chat interactions, showcasing the versatility of LLMs in software development.

In these applications, prompts serves as inputs to guide a generative AI model's output. They are the primary mechanism for communicating with the model \cite{Weber2024LargeLM}, making prompt design crucial to the effectiveness of these systems. Well-designed prompts ensure accurate and contextually relevant responses, while poorly formulated prompts can lead to ambiguity, misinterpretations, and incorrect output \cite{White2023APP}. Consequently, developers must iteratively refine their prompts in a process known as prompt engineering\cite{github_copilot_2021}, which is essential to achieve reliable and consistent model behavior.

%OLD:Beyond their functional importance, prompts represent a new type of software artifact, different from traditional components such as code, documentation, or comments \cite{liang2024prompts}. Prompts are natural language instructions embedded within software systems that directly influence the behavior of the integrated LLM. Unlike traditional software artifacts, prompts do not fit neatly into established categories and play a dual role at the intersection of functional specification and execution\cite{liang2024prompts}. 
Prompts, in addition to their functional role, represent a new type of software artifact distinct from traditional components like code, documentation, or comments \cite{liang2024prompts}. These natural language instructions are embedded within software systems, directly influencing the behavior of integrated LLMs. Unlike conventional software artifacts, prompts do not fit into established categories, serving a dual role between functional specification and execution \cite{liang2024prompts}. This unique role introduces new considerations for developers, who must manage and maintain prompts alongside conventional artifacts while recognizing their evolving nature and impact on system behavior.

% Despite the the growing production of LLM-integrated applications
%OLD:Given the distinct role of prompts as a new type of software artifact, there is limited understanding of how developers manage and engineer the prompts in LLM-integrated applications. Understanding these practices is crucial to gain insight into how developers interact with LLMs, the strategies they use to evolve prompts, and the challenges they face in prompt maintenance. Such insights can inform the development of improved tools and frameworks for prompt engineering, ultimately enhancing LLM integrations and ensuring better system reliability. Moreover, exploring prompt evolution practices can provide a deeper understanding of how prompts relate to other software artifacts and their impact on overall software quality and development productivity.

As prompts emerge as a distinct software artifact, there is a limited understanding of how developers manage and engineer them within LLM-integrated applications. Investigating these practices is essential to understanding how developers interact with LLMs, refine prompts, and address challenges in prompt maintenance. Insights into these practices can guide the creation of better tools and frameworks for prompt engineering, improving LLM integrations and system reliability. Additionally, studying prompt evolution can reveal how prompts interact with other software artifacts, influencing software quality and productivity.

%OLD: In this study, we present a comprehensive empirical analysis of the prompt evolution of LLM-integrated software development. By examining 1,262 prompt changes across 243 real-world software repositories hosted on GitHub, we investigate key aspects of prompt maintenance: modification patterns throughout the development lifecycle, relative change frequency compared to code, documentation practices, and the impact of prompt changes on system behavior. To better understand prompt management in practice, we devised a series of research questions organized across four themes:

This study presents a comprehensive empirical analysis of prompt evolution in LLM-integrated software development. By analyzing 1,262 prompt changes across 243 real-world GitHub repositories, we explore key aspects of prompt maintenance, including modification patterns throughout the development lifecycle, change frequency relative to code, documentation practices, and the impact of prompt changes on system behavior. To deepen our understanding of prompt management, we developed a series of research questions structured around four main themes:

\noindent\textbf{Theme 1: Prompt Change Types and Patterns}\\
\textit{Understanding the nature and evolution of prompt changes}
\begin{itemize}
    \item \textbf{RQ1:} What are the categories of changes made to prompts?
    \item \textbf{RQ2:} Which prompt components are changed, and how are they changed?
    \item \textbf{RQ3:} Is there a pattern of change in the prompts with multiple changes?
\end{itemize}

%We found that prompt evolution follows a gradual and iterative pattern, rather than undergoing abrupt changes. The most frequently modified components are the instructions and output format.
% Additionally, in prompts with multiple modifications, the most common pattern is the co-occurrence of changes to both instructions and demonstration examples.

\noindent\textbf{Theme 2: Prompt-Code Change Patterns}\\
\textit{Examining prompt changes in the broader context of software development}
\begin{itemize}
    \item \textbf{RQ4:} What is the software maintenance activity of prompt changing commits?
    \item \textbf{RQ5:} What is the distribution of prompt change types across different types of software maintenance activity? 
\end{itemize}

%Our analysis demonstrates that prompt modifications occur significantly less frequently than code changes. Moreover, prompt changes are distributed across different types of code modifications, however, they are most prevalent in feature development commits.

\noindent\textbf{Theme 3: Prompt Change Documentation and Intent}\\
\textit{Evaluating the documentation practices for prompt changes}
\begin{itemize}
    \item \textbf{RQ6:} How frequently do developers document prompt changes in the commit message?
    \item \textbf{RQ7:} What are the intentions behind the prompt changes, as documented in the commit message?
\end{itemize}

%We discovered significant gaps in prompt change documentation, with the majority of modifications lacking explicit documentation in commit messages. Among documented changes, the descriptions tend to be overly abstract and general.

\noindent\textbf{Theme 4: Prompt Change Impact and Effectiveness}\\
\textit{Analyzing the impact of prompt changes}
\begin{itemize}
    \item \textbf{RQ8:} How do prompt changes affect logical consistency in the prompt?
    \item \textbf{RQ9:} Does the intended prompt change align with the actual outcome when executed by the LLM?
\end{itemize}

%Our findings show that prompt changes can introduce inconsistencies in prompts, where different parts of the prompt contain logically conflicting statements. Second, our result from prompt execution reveals three distinct patterns: (1) aligned changes, where the LLM's response matches the intended modification, (2) misaligned changes, where the response deviates from expectations, and (3) resistant changes, where the output remains unchanged regardless of the prompt change.

% Our findings show that prompt changes can introduce inconsistencies in prompts. In addition, our empirical analysis of prompt execution reveals three distinct patterns: aligned changes, misaligned changes, and resistant changes.

Building on our analysis, we aim to provide a comprehensive understanding of prompt management in LLM-integrated software. Our findings contribute to the development of tools and methodologies that support effective prompt engineering and integration in real-world projects. Specifically, our contributions are:

\begin{itemize}
\item First empirical study analyzing prompt evolution in open source LLM-integrated applications.
\item Insights into prompt change patterns, frequencies, and correlation with code modifications to help develop better tools for managing prompt evolution.
\item Identification of documentation gaps in prompt changes, highlighting the need for better practices to improve traceability and collaboration.
\item Insights into the impact of prompt changes, revealing inconsistencies introduced by prompt change and instances where LLM responses deviate from intended outcomes, highlighting the need for robust testing and validation.
\end{itemize}
\section{Background}
\label{sec:bg}

\subsection{Developer-Written Prompts}
Developer-written prompts are typically embedded directly in software code, and are programmatic texts stored within variables, functions, or configuration settings to guide the behavior of integrated LLMs\cite{pister2024promptset}. They are often crafted as static but parameterized strings, incorporating placeholders to be dynamically populated during runtime.  These prompts serve specific single-use purposes like generating SQL queries, translating text, or providing responses to user input. Figure \ref{fig:dev-p} illustrates an example of a developer-written prompt extracted from the \textit{pyspark-ai/pyspark-ai} repository, featuring variables such as \texttt{columns}, \texttt{query}, and \texttt{search\_results}. These variables are dynamically populated at runtime, either through internal system configuration (columns), user inputs (query), or external data sources (search\_results). 
Similar to other software artifacts,
prompts require ongoing maintenance to reflect functionality changes and ensure quality. This work focuses on examining the evolution of developer-written prompts within open-source software repositories.

%OLD:Similar to other software artifacts, they require maintenance and evolve over time, reflecting changes in software functionality to ensure continued quality and alignment with intended behaviors. In our work, we focus specifically on developer-written prompts, examining their evolution within open-source software repositories.

% Developer-written prompts, embedded within software repositories, are programmatic texts that guide interactions with LLMs\cite{pister2024promptset}. These prompts often appear as static strings or dynamic templates that incorporate variables for adaptability during execution. An example of this prompt is shown in Fig \ref{fig:dev-p}. Unlike user-generated prompts in conversational AI systems, these prompts serve specific, functional purposes within applications, such as generating responses, summarizing texts, or handling system states. They are engineered components that follow coding standards, undergo version control, and evolve with the software codebase. In our work, we focus specifically on developer-written prompts, examining their evolution within real-world software repositories.

% \begin{figure}[htp]
%     \centering
%     \includegraphics[width=\columnwidth]{figures/dev-prompt.png}
%     \caption{Example of developer-written prompt}
%     \label{fig:dev-p}
% \end{figure}

\begin{figure}[htp]
    \centering
    \begin{lstlisting}[
        language=Python,
        basicstyle=\ttfamily\footnotesize,
        keywordstyle=\color{orange},
        stringstyle=\color{teal},
        commentstyle=\color{blue},
        backgroundcolor=\color{white},
        frame=single,
        xleftmargin=0.5em,
        xrightmargin=0.5em,
        breaklines=true,
        showstringspaces=false,
        columns=flexible
    ]
SEARCH_TEMPLATE = """
Given a Query and a list of Google Search Results,
return the link from a reputable website which
contains the data set to answer the
question. {columns}

Query:{query}

Google Search Results:
```
{search_results}
```

The answer MUST contain the url link only
"""

    \end{lstlisting}
    \caption{Example of developer-written prompt}
    \label{fig:dev-p}
\end{figure}

\subsection{Prompt Components}

A prompt often consists of multiple components, each contributing to eliciting a precise and relevant response from the model. Schulhoff et al. \cite{schulhoff2024prompt} identified six common components used by practitioners: (1) Directive—defines the desired task with instructions or questions; (2) Examples—illustrate the expected output, aiding in few-shot learning; (3) Output Formatting—specifies how the response should be structured (e.g., lists or JSON); (4) Style Specifications—dictates the tone and style of the response; (5) Role—assigns a perspective, such as an ``expert advisor'' or ``customer support agent''; and (6) Additional Information—provides extra context to ensure relevance and accuracy. Figure \ref{fig:prompt_components} shows an example illustrating these components.

For this study, we initially adopted the categorization defined by Schulhoff et al. However, based on our empirical evaluation of real-world prompts crafted by developers, we refined this categorization to better fit our dataset. Specifically, we merged two existing components into a single category and introduced a new component category to address elements not captured by the original categorization. Details of the refined categorization can be found in Section IV.

\section{Related Work}
\label{sec:rw}

\subsection{LLM-integrated Applications}

In recent years, there has been a significant surge in the development and deployment of LLM-integrated applications across various domains, including coding assistants \cite{github_copilot_2021, cursor_ai, tabnine}, healthcare aids \cite{nuance_dax, infermedica, abridge}, research tools \cite{elicit, semantic_scholar, research_rabbit}, and financial support systems \cite{wu2023bloomberggpt, ibm_watson_assistant, kasisto_kai}. This rise in LLM-integrated systems has introduced both unique challenges and opportunities for software engineering. Recent studies have examined how developers use tools like ChatGPT in open-source projects \cite{10555619, 10.1145/3674805.3690755, 10.1145/3597503.3608128, sauvola2024future}, explored challenges in maintaining LLM projects, and identified potential vulnerabilities in these systems \cite{cai2024demystifying, jiang2023identifying}. Researchers have also highlighted the distinct properties of these systems, such as prompt engineering and model orchestration, which require specialized software engineering practices \cite{10.1145/3663529.3663849}. Additionally, the iterative process of prompt programming has been identified as different from traditional software development, with developers facing unique challenges in crafting reliable prompts \cite{liang2024prompts}. Our research extends this by exploring the natural evolution of prompts in open-source software repositories, focusing on their integration and maintenance alongside other software artifacts.

\begin{figure}[t]
    \centering
    \includegraphics[width=0.80\columnwidth]{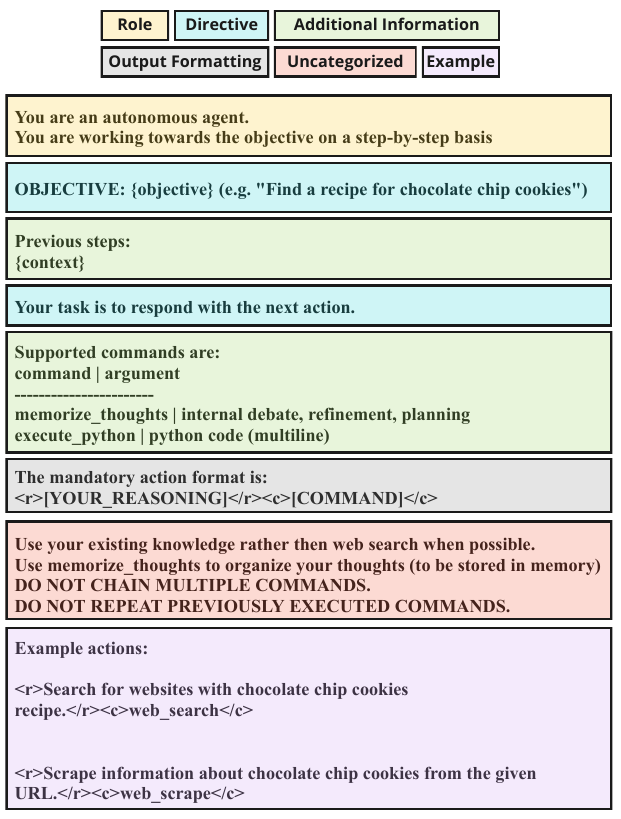}
    % \captionsetup{font=small}
    \caption{Example of prompt components introduced by Schulhoff et al. \cite{schulhoff2024prompt}}
    \label{fig:prompt_components}
\end{figure}

\subsection{Prompt Engineering and Evolution}
%OLD:Prompt engineering is the strategic crafting of prompts to elicit desired responses from LLMs without additional fine-tuning\cite{sahoo2024systematic}. Recent research has demonstrated its effectiveness across various software engineering tasks. For instance, Shin et al. \cite{shin2023prompt} showed that GPT-4, when guided by task-specific prompts, could match the performance of fine-tuned models in code translation. Similarly, Gao et al. \cite{gao2023makes} found that carefully selecting and ordering prompt examples significantly improved accuracy and consistency in code summarization and bug fixes.
Prompt engineering involves strategically crafting prompts to elicit desired responses from LLMs without fine-tuning \cite{sahoo2024systematic}. Recent studies have highlighted its effectiveness across various software engineering tasks. For example, Shin et al. \cite{shin2023prompt} demonstrated that GPT-4, guided by task-specific prompts, could match the performance of fine-tuned models in code translation. Similarly, Gao et al. \cite{gao2023makes} showed selecting and ordering prompt examples improved both accuracy and consistency in code summarization and bug fixes.

Automated approaches to prompt evolution have also gained traction. PromptBreeder \cite{fernando2023promptbreeder} evolve prompts through multiple generations using LLM-driven mutations, while SPELL \cite{li2023spell} optimizes prompts for coherence and quality through a black-box LLM mechanism. These approaches focus on automated optimization, yet our study investigates the natural evolution of prompts as developers within open-source software repositories iteratively refine them. 
Additionally, Desmond et al. \cite{desmond2024exploring} analyzed prompt engineering within a controlled internal platform called Enterprise. They categorized prompt edits and components to understand the use cases of prompt engineering in practice. However, their study was limited to categorizing prompt changes, while our research examines prompt evolution within a broader context of software development, including its relationship with code, documentation, and system maintenance.

\section{Methodology}
\label{sec:method}

In this section, we present our research methodology. As illustrated in Figure \ref{fig:overall-flow}, our approach consists of three main phases: (1) Data Preparation, (2) qualitative analysis, and (3) quantitative analysis. We provide detailed descriptions in the following subsections.

\subsection{Dataset Collection}

To examine the evolution of prompts in open source software repositories, it is essential to extract developer-defined prompts from these repositories. For this purpose, we utilized an existing dataset named PromptSet \cite{pister2024promptset}, which contains 61,448 unique prompts collected from 20,598 Python projects hosted on GitHub. This dataset provides information such as the repository name, file path within the repository containing prompts, and an array of raw prompt texts in the corresponding file. However, PromptSet has certain limitations. First, no quality assessment of this dataset is available in the literature, and the original authors made no effort to clean the data. Our manual inspection also revealed several issues, including entries where no prompts were collected for the specified file paths. Additionally, we observed that many prompts in the dataset are fragmented or incomplete. 
This is because developers often split the different components of a prompt, such as directives or examples, into separate variables or string literals and then concatenate them when sending the message to an LLM in an API call. As a result, PromptSet cannot capture the complete prompt and instead contains many fragmented prompts, making them difficult to analyze in isolation. Many of such prompts contain only a few words. Due to these limitations, we performed a data-cleaning phase before using PromptSet in our study.

\subsection{Data Cleaning and Filtering}
We started the data cleaning by removing empty prompts from the dataset. Next, we filtered out prompts that were not in English or contained non-ASCII characters. Finally, we eliminated incomplete and distorted prompts with fewer than 15 words. We decided to use the threshold of 15 words for two-fold reasons. Our manual analysis revealed that prompts with lengths less than 15 are usually less suitable for analysis and often contain prompts such as ``You are a helpful assistant.'' Moreover, research indicates that an average English sentence has around 15 to 20 words\cite{feng2010comparison}. After these steps, the dataset was reduced to 32,819 prompts from 18,692 repositories. Following the initial cleaning, we proceeded with repository filtering to focus on popular and actively maintained projects\cite{10.1145/2597073.2597074, pickerill2020phantom}. To achieve this, we applied the following criteria: repositories with at least 50 stars, a minimum of 10 contributors, and an active development period of at least 6 months. Applying these filters resulted in a set of 1,568 repositories.

\subsection{Prompt Change Collection}
After finalizing our list of repositories, we cloned each one and collected prompt changes using the Git commit history. Utilizing the file path attribute, we started from the initial commit that created the file containing prompts and traced all subsequent versions up to August 1st, 2024. For each version of a file, we used the PromptSet tool to extract the prompts. We then performed a pairwise comparison between consecutive prompt versions to identify any changes. When changes were detected, we recorded both the old and new versions of the prompt, together with the similarity score (calculated using the ROUGE-L \cite{lin2004rouge} similarity metric). After removing duplicate changes, we arrived at a final dataset of 1,262 unique prompt changes in 243 repositories.

\begin{figure*}[t]
    \centering
    \includegraphics[width=1.0\textwidth]{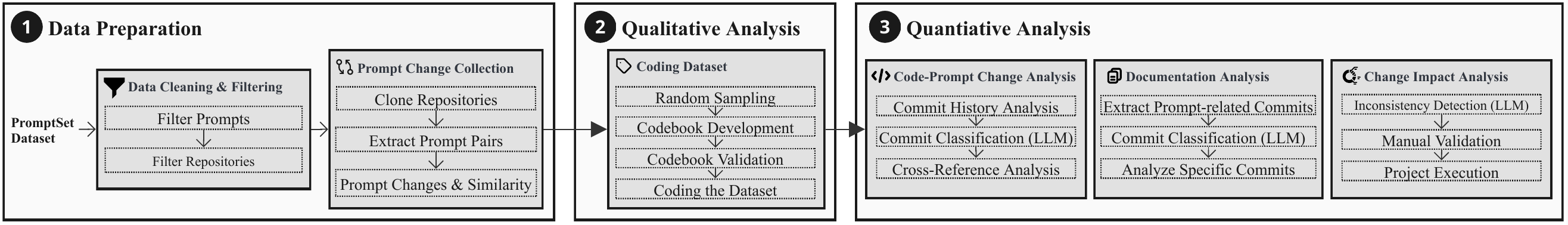}
    \captionsetup{font=small}
    \caption{Overall methodology}
    \label{fig:overall-flow}
\end{figure*}

\subsection{Qualitative Analysis}
To understand the types of prompt changes made to the prompts and the their modified components, we conducted a qualitative analysis. The analysis is done in two phases: codebook development, which involves creating a structured set of categories for prompt changes and their affected components, and qualitative coding, where these categories are applied to label each prompt change.
To enhance the coding process, we highlighted the differences between the old and new versions of each prompt and used the Label Studio platform \cite{labelstudio} for the labeling. In the first phase of codebook development, we randomly selected a sample of 225 from a total population of 1,263 (confidence level 90\%, margin of error 5\%). This sample size was determined to be statistically representative of the entire population. Two authors independently analyzed these prompt changes, adding initial codes and comments to capture observed changes in both structure and content. Following the independent coding phase, the authors came together in a negotiated agreement process, where they compared their findings, discussed differences, and refined their coding strategies \cite{corbin2015basics,strauss1998basics}. Through this collaboration, they reached a consensus on the final set of codes, resulting in a comprehensive codebook that detailed both the types of edits made to the prompts and the specific components edited. We will discuss the identified types in Section \ref{sec:results}.
% , as categorized in Table \ref{tab:prompt_categories}.

To further validate the codebook, the authors conducted an additional round of coding to assess inter-rater reliability \cite{landis1977measurement}. A new sample of 65 prompt changes was independently coded by both authors using the finalized codebook. Given the complexity of prompt changes, where multiple codes could apply to each change and the presence of multiple categories, \textit{Jaccard Index Averaging}\cite{real1996probabilistic} was employed to calculate inter-rater reliability.

The \textit{Jaccard Index} $J_i$ for each prompt change $i$ was calculated as:

\begin{equation}
J_i = \frac{|C_{A_i} \cap C_{B_i}|}{|C_{A_i} \cup C_{B_i}|}
\end{equation}

where $C_{A_i}$ and $C_{B_i}$ represent the sets of codes assigned by \textit{Author A} and \textit{Author B} for prompt change $i$, respectively. The \textit{intersection} $|C_{A_i} \cap C_{B_i}|$ captures the number of codes both raters agreed upon, while the \textit{union} $|C_{A_i} \cup C_{B_i}|$ accounts for all unique codes assigned by either rater.

To obtain an overall measure of inter-rater reliability, we averaged the Jaccard Index across all prompt changes:

\begin{equation}
J_{avg} = \frac{1}{n} \sum_{i=1}^{n} J_i
\end{equation}

where $n$ represents the total number of prompt changes analyzed. Using this method, the final inter-rater agreement reached 65\%, a level that was acceptable given the complexities of labeling prompt components, various edit types, and the overall number of categories in the codebook. This level of agreement reflects the challenges inherent in multi-label coding. After reaching an agreement on the codebook and establishing inter-rater reliability, the authors proceeded with coding the remaining dataset of 1,262 prompt changes. They divided the dataset, with each author independently coding their assigned subset.

\subsection{Prompt-Code Change Analysis}

We analyzed the frequency and context of prompt changes concerning code changes. The process involved the following steps:
% \subsubsection{Commit History Analysis} We used the Git commit history of each repository to determine the frequency of prompt changes compared to code changes. We started from the first commit, where prompts were introduced to the project, and for each subsequent commit, we identified whether it involved a prompt change, a code change, or both. This allowed us to measure and compare the occurrence rates of prompt and code modifications across the entire dataset.

\subsubsection{Commit Classification into Software Maintenance Activities}
To analyze the relationship between prompt changes and software maintenance activities we classified commits into specific maintenance categories. For classification, we followed an approach similar to prior works \cite{imani2024context,li2024only}, which represent the state-of-the-art with an accuracy of 51\%, leveraging an LLM to analyze the commit message and code diff for each commit to assign a software maintenance activity. Specifically, we utilized a 4-bit quantized Llama3 70B model \cite{dubey2024llama} to categorize commits into three primary maintenance activities: feature addition, bug fixing, and refactoring. Once categorized, we analyzed the distribution of prompt changes across these different commit types to understand the relationship between prompt modifications and specific software maintenance activities.

% \subsubsection{Commit Classification to Software Maintenance Activities}

% For commit classification into software maintenance activity similar to the prior works \cite{imani2024context,li2024only} we leveraged an LLMs and prompting it with the commit message and code diff of each commit. Specifically, we used a 4-bit quantized LLama3 70B \cite{dubey2024llama}.  Commits were categorized as feature, bug fix, or refactoring. We then examined how prompt changes were distributed across these commit types to determine the co-occurrence of prompt modifications with specific kinds of software maintenance activity.

\subsubsection{Analyzing the Distribution of Prompt Change Types Across Software Maintenance Activities} After classifying each commit into a specific software maintenance activity we cross-referenced the identified prompt change categories with these classifications. This mapping allowed us to determine which types of prompt changes were most frequently associated with each kind of software maintenance activity.
% \hl{what correlation analysis was used? mention}

\subsection{Prompt Change Documentation and Intent Analysis}

To address the research questions under Theme 3, we focused on understanding how prompt changes are documented in the commit message and the underlying reasons for these modifications. The analysis involved the following steps:

\begin{figure*}[b!]
    \centering
    \includegraphics[width=1.0\textwidth]{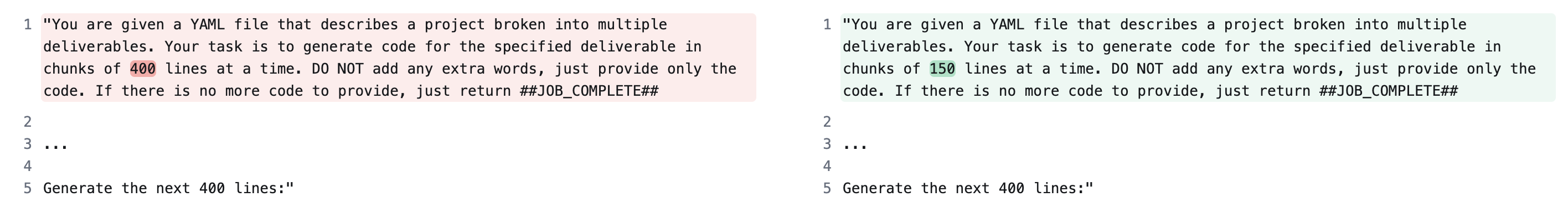}
    \captionsetup{font=small}
    \caption{Inconsistency in prompt statements introduced by directive modification (murchie85/gpt\_automate)}
    \label{fig:inc}
\end{figure*}

\subsubsection{Prompt-related Commit Extraction} 

To effectively understand the documentation practices for prompt changes, we extracted commit messages that indicated modifications to prompts from the selected repositories. For this extraction, we again leveraged LLMs, which are well-suited for textual analysis. Specifically, we used a 4-bit quantized LLama3 70B \cite{dubey2024llama} to identify commit messages referencing prompt changes. We employed a few-shot prompting approach, where we crafted prompts with example commit messages involving prompt modifications and instructed the model to locate similar messages across the dataset.

\subsubsection{Related Commit Classification} After extracting 204 related commit messages, we manually categorize them into \textbf{General} and \textbf{Specific} groups, based on the level of detail provided. General messages contained limited information about the purpose of the prompt change. An example of these messages are ``Update prompts'' or ``Tweak prompts.'' Specific messages provided more detailed insights into what was changed and in some cases the reason behind this change, such as ``Modified prompt for GPT3.5 to more reliably provide questions in the correct format''.
% To ensure accuracy, we manually reviewed the classification of commit messages, correcting any false positives or misclassifications.
% \hl{we need to report numbers for this}

\subsubsection{Analysis of Specific Commit Messages} For commit messages in the Specific category, we conducted a qualitative analysis to uncover the underlying reasons for prompt changes. Following standard open coding practices, two authors independently performed coding of the intention of prompt change in commit messages. After independent coding, the authors met to compare their codes, resolve disagreements through discussion, and iteratively refine the coding scheme until reaching a consensus on the classification of changes \cite{corbin2015basics,strauss1998basics}. 
% We categorized these reasons into types such as hallucination handling, template updates, and output modifications. This categorization helped us to understand the motivations behind the prompt modifications in more detail. 

\subsection{Prompt Change Impact and Effectiveness Analysis}
To explore the impact of prompt changes on system behavior and assess the effectiveness of these modifications, we conducted an analysis using both automated and manual methods. We aimed to understand how prompt changes affected prompt consistency and whether the intended changes aligned with the LLM's response.

\subsubsection{Inconsistency Detection} To detect inconsistencies introduced by prompt changes, we leveraged an LLM using a chain-of-thought prompting approach \cite{wei2022chain, zhang2024detecting, yang2024sifid}. Specifically, we asked the model first to read both the old and new prompts, then identify the modified parts, and lastly, for each change, determine whether this modification introduced inconsistencies within the prompt statements. Inconsistency was defined as occurring when the changed part of the prompt introduced conflicting information or contradicted the unchanged parts of the prompt. This automated process reduced the number of potentially inconsistent prompt changes to 205 instances, which were then manually reviewed by two authors. During manual analysis, we verified the inconsistencies and evaluated the reasons provided by the LLM. An example of inconsistency can be found in Figure \ref{fig:inc}. The modified prompt asks to generate code in chunks of 150 lines, but still requests to generate the next 400 lines, which is a conflicting instruction.

\subsubsection{Impact Analysis of Prompt Changes on LLM Responses} To determine whether the intended prompt changes aligned with the actual outcomes when executed by the LLM, we focused on projects where prompt modifications were explicitly documented in commit messages. This focus allowed us to understand the motivations behind each modification better, providing critical context for evaluating each change's intended purpose and impact. That resulted in a total of 101 projects. We manually analyzed each of these projects to replicate prompt execution and assess the impact of the changes on the LLM's response. This process presented several challenges that should be explored further and have implications for future research. Many projects lacked test cases that specifically targeted the parts of the program interacting with the LLM, highlighting the novelty and complexity of evaluating LLM components within software development. Additionally, prompts often included variables defined at runtime—such as through user inputs or dependencies like databases and external APIs—making it challenging to reconstruct complete prompts without specific examples or context. These challenges suggest the need for better tooling and methodologies for assessing LLM prompt changes. Out of the 101 projects we analyzed, we successfully executed test cases from 7 projects that provided suitable test cases or example inputs. These projects enabled us to evaluate the LLM responses before and after prompt modifications and determine whether the responses aligned with the intended changes.

\begin{mdframed}[roundcorner=10pt]
\textbf{Observation 1}: Prompt changes primarily target specific components through additions and modifications, while changes affecting the overall prompt structure or presentation occur less frequently and focus mainly on clarity improvements.
\end{mdframed}
\section{Results}
\label{sec:results}

In this section, we present the results of our study with respect to our study themes and their underlying RQs.

\subsection{Theme 1: Prompt Change Types and Patterns}

To address the research questions related to this theme, we present the outcomes of our qualitative analysis process.

% \\ \textit{Understanding the nature and evolution of prompt changes}

\begin{table*}[t!]
\captionsetup{font=small}
\caption{Categories of Prompt Changes, Definitions, and Frequencies}
\label{table:change-type-frequency}
\resizebox{\textwidth}{!}{%
\begin{tabularx}{\textwidth}{ccXcc}
\toprule
\textbf{Change Category} & \textbf{Change Type} & \textbf{Definition} & \textbf{Frequency} & \textbf{Percentage} \\
\midrule
\multirow{3}{*}{Component-dependent} & Addition & Introduction of new component to an existing prompt, such as new instructions or examples & 440 & 30.1\% \\
 & Modification & Changes that alter the semantic meaning or behavior specification of existing prompt components & 373 & 25.5\% \\
& Removal & Elimination of existing prompt components & 129 & 8.8\% \\
\midrule
\multirow{6}{*}{Component-independent} & Rephrase & Changes that preserve the original semantic meaning while altering the way it is expressed & 254 & 17.4\% \\
& Formatting & Changes to the prompt's template, whitespace, or other aesthetic components & 116 & 7.9\% \\
& Correction & Fixes for spelling or grammar errors & 71 & 4.9\% \\
& Generalization & Replacement of part of a prompt with variables or placeholders & 34 & 2.3\% \\
& Restructure & Reorganization of existing prompt components without adding or removing content & 29 & 2.0\% \\
& Uncategorized & Changes that were unique or ambiguous in nature & 17 & 1.1\% \\
\midrule
\textbf{Total} & & & \textbf{1,463} & \textbf{100\%} \\
\bottomrule
\end{tabularx}
}
\end{table*}

\begin{figure}[t]
    \centering
    \includegraphics[width=0.75\columnwidth]{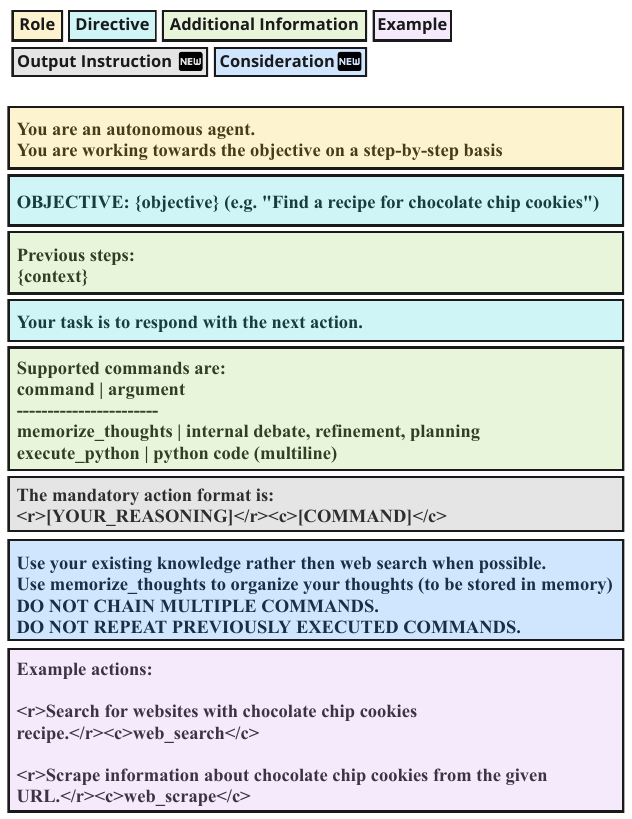}
    \caption{Example of refined prompt components}
    \label{fig:enhanced_prompt_components}
\end{figure}

\begin{figure}[t]
  \centering
  \includegraphics[width=0.82\linewidth]{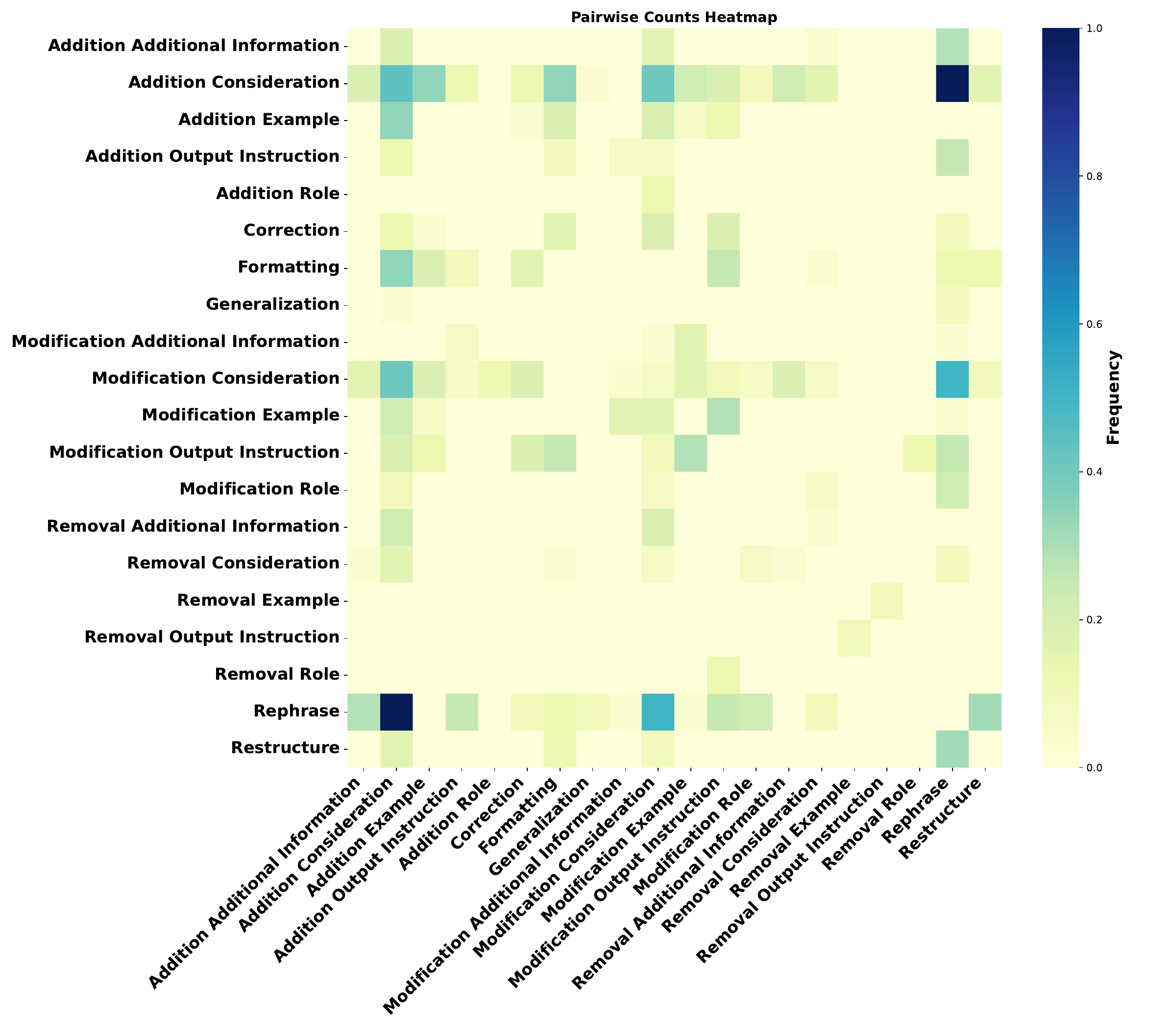}
  \captionsetup{font=small}
  \caption{Proportional Distribution of Changes Across Prompt Components}
  \label{fig:heatmap}
\end{figure}

\subsection*{\textbf{RQ1:} What are the categories of changes made to prompts?}

Our qualitative analysis of 1,262 changed prompts resulted in 1,463 distinct prompt changes. 
This difference is due to observing multiple changes per changed prompt on the majority of records.
% This difference is because developers often make multiple edits at the same time on a prompt. 
We will discuss the co-occurrence of these changes and their patterns in detail in RQ3.
% Through our analysis of prompt changes,
Other than 1.1\% of prompt changes which we could not categorize into any change type due to their ambiguity, we identified eight distinct types of changes that developers make when evolving prompts (Table \ref{table:change-type-frequency}). These changes are categorized into component-dependent and component-independent modifications based on whether they target specific prompt components or affect the prompt's overall structure and presentation. 

Component-dependent changes, accounting for 64.4\% of all changes, directly affect specific components within prompts. 
Additions (30.1\%) emerge as the most frequent type, indicating that initial prompts often require expansion over time. 
% Modifications (25.5\%) represent changes that alter the semantic meaning or behavior specification of existing components. 
While removals are also component-dependent, they occur less frequently (8.8\%), suggesting that developers typically expand and enhance their prompts rather than reduce them. 

Component-independent changes, comprising 34.5\% of modifications, affect the prompt's overall structure or presentation without targeting specific components. Rephrasing (17.4\%) is the most common in this category, following two main approaches: simplifying complex instructions or enhancing clarity through better expression. Formatting changes (7.9\%) focus on template and whitespace modifications, while corrections (4.9\%) address basic spelling and grammar issues. Generalization (2.3\%) and restructuring (2.0\%) are the least frequent, indicating that developers rarely make broad structural changes once a prompt's organization is established.

\subsection*{\textbf{RQ2:} Which prompt components are changed, and how are they changed?}

As discussed in Section \ref{sec:method}, we began coding using the six prompt components introduced in Section \ref{sec:bg}. However, we found that \textit{Output Formatting} and \textit{Style Specifications} frequently overlapped in the initial sample, leading us to consolidate them into a single category named \textit{Output Instruction}. Additionally, we identified prompt elements that did not fit into any of the six original prompt components. These elements represent commands that guide the language model in performing assigned tasks, characterized by either specific instructions or constraints, which we classified as \textit{Consideration}. Consequently, our refined categorization comprises six distinct components, as presented in Table \ref{table:component-frequency}. This updated component categorization better reflects the strategies employed by developers in prompt construction. An example illustrating the use of these components is depicted in Figure~\ref{fig:enhanced_prompt_components}.
Among these components, \textit{Consideration} was the most frequently modified, accounting for 60.4\% of all prompt changes. This high frequency suggests that developers prioritize refining the instructions or constraints that guide the behavior of the LLM. Changes to \textit{Consideration} were primarily additions and modifications, reflecting developers' ongoing efforts to adjust the model’s behavior to evolving requirements. The high number of additions implies a continuous effort to provide more detailed instructions, enhancing the comprehensiveness of the guidance provided to the LLM.
% To gain further insights, we analyzed the \textit{Consideration} changes in more depth. Our findings reveal that in 502 instances (89\% of all changes), developers modified \textit{Consideration} to change the nature of the guiding task assigned to the model. In 58 cases (10\%), developers aimed to clarify the \textit{Consideration}—adding more explicit details or explanations. Lastly, in 9 instances (1\%), changes emphasized existing \textit{Considerations}, suggesting that developers sought to stress particular guidelines to ensure the model adhered strictly to these instructions. This detailed breakdown highlights that the primary purpose behind modifying \textit{Considerations} is to better control and guide the LLM's behavior, ensuring it is precisely aligned with the developers’ goals.

The \textit{Additional Information} component was the second most frequently changed, often through additions intended to provide further contextual details and enhance the quality of the LLM’s responses. Adding extra \textit{Additional Information} has proven effective in previous LLM-based studies \cite{imani2024context}.
Removing parts of this component, though less frequent, were typically aimed at simplifying the provided context or moving the context to another prompt.

\textit{Output Instruction} was the third most frequently changed component, with developers often refining the formatting requirements to ensure consistency in the LLM-generated responses. These modifications suggest a focus on changing the format, structure or length of the outputs.

Changes to the \textit{Example} component account for approximately 8\% of all changes. Given the significant role of examples in prompting techniques such as In-context Learning \cite{min2022rethinking}, further research is recommended to explore the reasons behind developers’ modifications to this component and the anticipated outcomes of such changes.

The \textit{Role} component was the second least changed, appearing in approximately 3\% of the changes. This low frequency aligns with prior research \cite{hu2024quantifying}, which suggests that altering the model’s role or persona has minimal impact on output quality from developers' perspective. Once a role is defined, it tends to remain stable, indicating that developers do not frequently change LLM's role or persona.

Finally, changes to the \textit{Directive} component were the least frequent. Directives define the core task or high-level objective of the LLM, and the rarity of changes to this component indicates that developers primarily establish the fundamental purpose at the outset, with subsequent modifications focusing on refining other aspects of the prompt.

\begin{mdframed}[roundcorner=10pt]
\textbf{Observation 2}: Developers primarily focus on refining \textit{Consideration} that guides the model behavior, followed by adjustments to contextual information and output formatting, while core components stay stable.
\end{mdframed}

\begin{table*}[b!]
\captionsetup{font=small}
\caption{Frequency of Edits on Each Prompt Component}
\label{table:component-frequency}
\resizebox{\textwidth}{!}{%
\begin{tabularx}{\textwidth}{X|ccc|cc}
\toprule
\textbf{Prompt Component} & \textbf{Addition (\%)} & \textbf{Modification (\%)} & \textbf{Removal (\%)} &  \textbf{Change Frequency} &  \textbf{Change Percentage} \\
\midrule
Consideration & 51.1	& 35.7 & 13.2 & 569 & 60.40\% \\
Additional Information & 36.0 & 27.9 & 36.0 & 154 & 16.35\% \\
Output Instruction & 25.9 & 71.4 & 2.7 & 112 & 11.89\% \\
Example & 44.00 & 46.7 & 9.3 & 75 & 7.96\% \\
Role & 13.3 & 73.3 & 13.3 & 30 & 3.18\% \\
Directive & - & 100.0 & - & 2 & 0.21\%
\end{tabularx}
}
\end{table*}

\subsection*{\textbf{RQ3:} Is there a pattern of change in the prompts with multiple changes?}
Developers often make multiple types of changes simultaneously when modifying a prompt. We are interested in understanding whether there are patterns in these simultaneous changes, as such patterns could reveal common prompt engineering strategies. To better analyze these co-occurring changes, we produced a heatmap to visualize the proportional frequency of co-occurring prompt changes. Figure~\ref{fig:heatmap} presents this heatmap.

Our analysis reveals several consistent patterns. The most frequent combination is the additions to the Consideration component paired with rephrasing, followed by modification of Consideration and rephrasing. They are indicated by the darkest colors in their respective rows and columns in the heatmap. This pattern suggests that when developers add or modify instructions and constraints, they often simultaneously refine the clarity of expression.

We also observed frequent pairings between adding Considerations and either adding Examples or making Formatting changes. These combinations suggest that developers often support additions to  the \textit{Consideration} component with demonstrative examples or adjusting the prompt's structure. Notably, certain changes rarely or never co-occur. For instance, removals of different components rarely happen at the same time.

\begin{mdframed}[roundcorner=10pt]
\textbf{Observation 3}: Additions and modifications to Consideration are typically accompanied by rephrasing changes, while formatting and example additions often support new Consideration elements.
\end{mdframed}

% To answer this research question, we created a heatmap to visualize the proportional frequency of co-occurring prompt changes. Figure \ref{fig:heatmap} presents this heatmap. X and Y are the most frequently co-occurring prompt changes, as indicated by the darkest colors in their respective rows and columns. In contrast, Z and GAMMA are isolated prompt changes that did not co-occur with any other types of prompt changes.

\subsection{Theme 2: Prompt-Code Change Patterns}
% \hl{Isn't figure 7 supposed to be referenced? AI: Added them to RQ5 answer}
To address the research questions within this theme, we analyzed the frequency and context of prompt changes in relation to code modifications. Below, we present the findings of this analysis to answer the research questions.

% \subsection*{\textbf{RQ4:} How do prompt change frequencies compare to code changes?}

% TBD after Mahan talks to Iftekhar

\subsection*{\textbf{RQ4:} What is the software maintenance activity of prompt changing commits?}

To understand the context in which developers modify prompts, we analyzed the software maintenance activities associated with prompt-changing commits. As discussed in Section \ref{sec:method}, we classified commits into different maintenance activities based on their commit messages and code changes. 
Our analysis reveals that 
% According to Figure \ref{fig:rq5},
prompt changes predominantly occur during feature development, accounting for 59.7\% of all prompt-changing commits. This suggests that prompts are primarily evolved as part of new functionality implementation rather than maintenance tasks. Bug fixing activities constitute 18.9\% of prompt changes, followed by refactoring activities that account for 14.5\% of the changes.

\begin{mdframed}[roundcorner=10pt]
\textbf{Observation 4}: Prompt changes are predominantly associated with feature development, while bug fixes and refactoring play secondary roles in prompt evolution.
\end{mdframed}

% According to our commit classification results, the majority of prompt changes (360) occurred in feature addition commits. A total of 243 prompt-changing commits were classified as Refactoring commits. Bug fix commits represented the smallest number of commits incorporating prompt changes, with 218 instances. The remaining 235 commits did not fall into any of these specific commit categories.

\subsection*{\textbf{RQ5:} What is the distribution of prompt change types across different types of software maintenance activity?}

Figures \ref{fig:rq5} and \ref{fig:rq5-2} show the distribution of component-specific and component-independent changes across the software maintenance activities.
Our analysis reveals distinct patterns in how different types of prompt changes are distributed across maintenance activities. In feature development commits, we observe a strong focus on expanding and clarifying prompt content, with Addition Consideration and rephrasing being the most frequent changes. This suggests that when implementing new features, developers prioritize adding new instructions and ensuring their clarity.

Bug fixing activities show a different pattern, with Modification Consideration being the most common change type, followed by additions and rephrasing. Notably, corrections appear more frequently in bug fixes compared to other maintenance activities. 

Refactoring commits exhibit a unique distribution, with rephrasing being the most frequent change type. The prominence of rephrasing in refactoring commits suggests that developers focus on improving prompt clarity and expression without necessarily changing the underlying functionality.

\begin{mdframed}[roundcorner=10pt]
\textbf{Observation 5}: Feature development emphasizes adding new instructions, bug fixes focus on modifying existing behavior, and refactoring prioritizes clarity improvements through rephrasing.
\end{mdframed}

\begin{figure}[t]
    \centering
    \includegraphics[width=0.85\columnwidth]{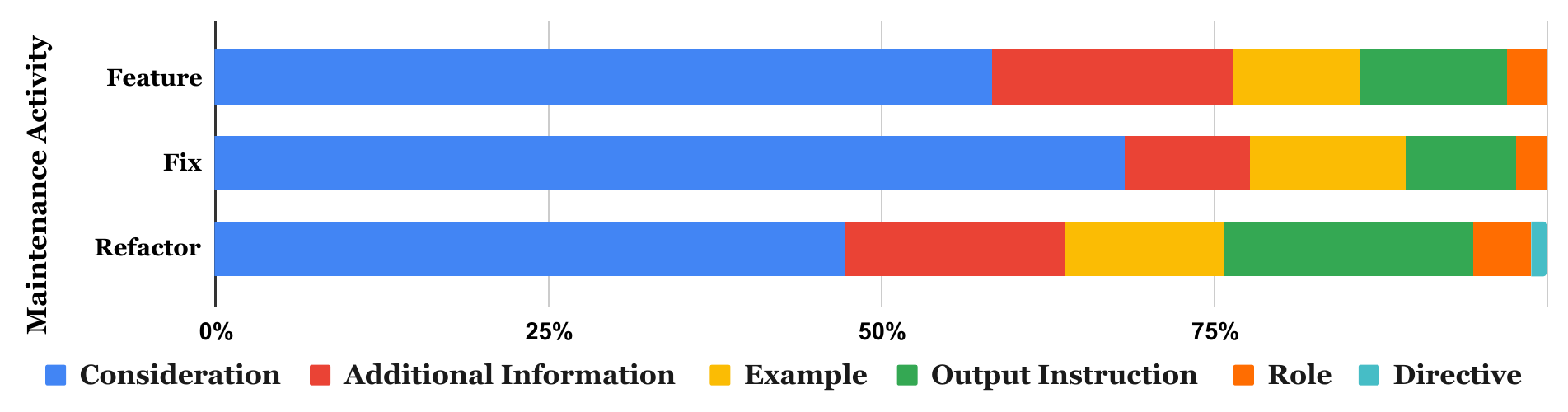}
    \captionsetup{font=small}
    \caption{Distribution of prompt component changes over software maintenance activity types}
    \label{fig:rq5}
\end{figure}

\begin{figure}[t]
    \centering
    \includegraphics[width=0.85\columnwidth]{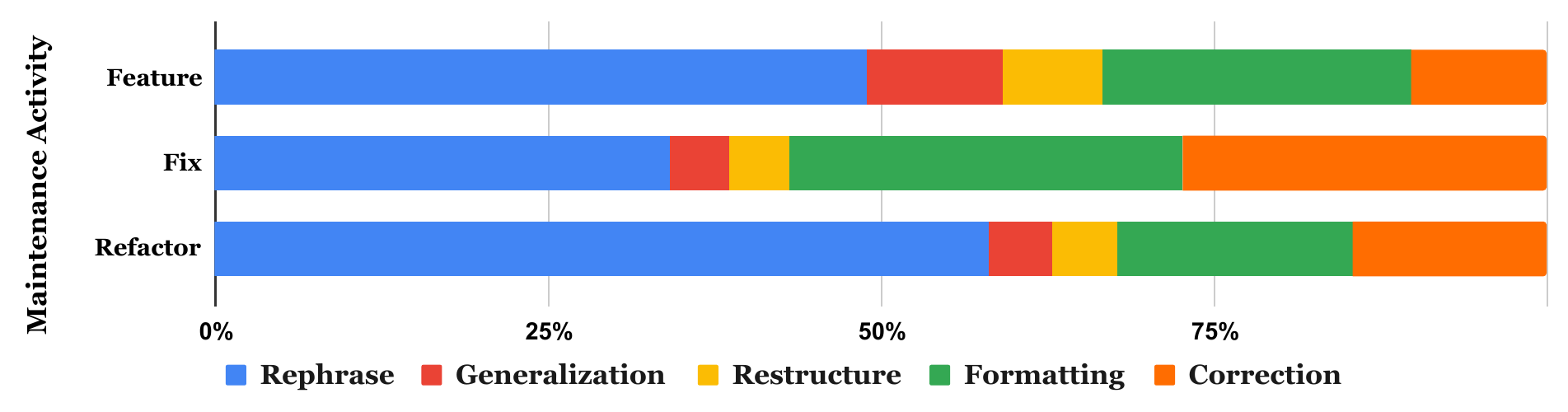}
    \captionsetup{font=small}
    \caption{Distribution of component-independent changes over software maintenance activity types}
    \label{fig:rq5-2}
\end{figure}

\subsection{Theme 3: Prompt Change Documentation and Intent}
\subsection*{\textbf{RQ6:} How frequently do developers document prompt changes in the commit message?}

To understand how developers document their prompt modifications, we analyzed the total of 931 commit messages in our dataset. We identified 204 commits (21.9\%) that explicitly referenced prompt changes, indicating that developers often do not specifically mention prompt changes in their commit messages.

Among the commits that do reference prompt changes, we observed two distinct documentation patterns. The majority (142 commits, 69.6\%) contain only general descriptions, using broad terms like ``Improve prompts'', ``Update prompts'', or ``Tweak prompts''. These general messages provide limited insight into what was changed or why the change was made. The remaining commits (62, 30.4\%) include specific details about the changes, describing either the nature of the modification, its purpose, or both. For example, messages like ``Fix context position in Llama2 QA prompt'' or ``Improves prompts with more focus on task completion'' provide clearer documentation of the changes and their intended effects.

% This lack of comprehensive documentation is particularly concerning given the increasing integration of LLMs in software systems. Prior research has shown that detailed documentation of code changes is crucial for software maintenance and debugging \cite{10.1145/2393596.2393656, tryggeseth1997report}, with undocumented changes often leading to increased maintenance difficulties \cite{tryggeseth1997report}. As prompts emerge as a new type of software artifact with direct impact on system behavior, similar documentation principles should apply. However, the current state of documentation practices appears insufficient. The low frequency and lack of specificity in prompt change documentation could hinder maintainability, complicate debugging efforts, and make it difficult for development teams to track the evolution of their LLM interactions. These findings suggest a clear need for establishing better documentation practices specifically tailored to prompt engineering.

The current state of prompt documentation practices is concerning, with only 21.9\% of prompt changes explicitly documented in commit messages, most of which lack detail. Unlike traditional code documentation that focuses on implementation details, prompt documentation must also capture intended behavior and the impact on LLM responses. Prior research has shown that undocumented changes can hinder maintenance and debugging \cite{10.1145/2393596.2393656, tryggeseth1997report}. Given the unique role of prompts in shaping LLM behavior, clear and detailed documentation is essential to ensure maintainability and traceability in LLM-integrated systems. Practitioners should prioritize comprehensive documentation of prompt changes, while researchers have the opportunity to extend existing automated commit message generation tools \cite{liu2020atom, nie2021coregen, dong2022fira} to include prompt modifications, effectively capturing both the nature of changes and their intended effects.

\begin{mdframed}[roundcorner=10pt]
\textbf{Observation 6}: Only 21.9\% of prompt changes are documented in commit messages, with most providing abstract and general descriptions instead of specific details about the prompt change.
\end{mdframed}

\subsection*{\textbf{RQ7:} What are the intentions behind the prompt changes, as documented in the commit message?}

% \hl{we should have a table for this AI: fixed}
To understand the developer's motivations for changing prompts, we qualitatively analyzed 62 commits in the \textbf{Specific} category. 
Table \ref{table:prompt-change-types} presents the results of our analysis.
We identified seven distinct categories: task-specific modifications, typo and grammar corrections, output modifications, template updates, technology adaptations, reformatting, and hallucination handling. 
% as summarized in\textit{ Table \ref{table:prompt-change-types}.}

Task-specific modifications exhibit three main patterns in our analysis. First, developers frequently focus on improving data processing capabilities and handling complex operations, as seen in commits like ``Avoid overfit in prompt transformation'' and ``Improve prompt to help query correct column.'' Second, they enhance context accuracy, demonstrated by commits like ``Passing missing context into prompt.'' Third, they clarify specific behaviors, exemplified by commit messages such as ``Add explanation of when to use SUM vs. COUNT.''

Basic maintenance activities comprise the second major group of changes. These include typo fixes and grammar corrections (11.3\%), improving prompt clarity, and output formatting modifications (8.1\%), addressing how LLMs structure their responses. For example, commits like ``Fix typo in moderator prompt'' handle basic text corrections, while ``Modify verify prompt to exclude quotes around function definition'' and ``Add additional prompt to return tabular data as html table'' demonstrate efforts to control output format. Technology adaptations (changing prompts due to changes in underlying models, e.g., ``adjust prompt for gpt-4-turbo''), template updates, reformatting, and hallucination handling collectively constitute 16\% of the change intentions.

\begin{mdframed}[roundcorner=10pt]
\textbf{Observation 7}: Documented prompt changes primarily aim to improve task-specific behavior rather than addressing broader prompt engineering concerns such as hallucination control or model adaptations.
\end{mdframed}

\begin{table}[t!]
\caption{Categories of Changes in Prompt Modifications}
\label{table:prompt-change-types}
\centering
\begin{tabular}{p{0.7\columnwidth}r}
\toprule
\textbf{Category} & \textbf{Count} \\
\midrule
Task Specific Change & 40 \\
Typo Fix \& Grammar Correction & 7 \\
Output Modification & 5 \\
Template Update & 3 \\
Technology Change/Adaptation & 3 \\
Reformat (moving parts of the prompt) & 2 \\
Handle Hallucination & 2 \\
\midrule
\textbf{Total} & \textbf{62} \\
\bottomrule
\end{tabular}
\end{table}

\begin{table}[t!]
\caption{Categories of Inconsistencies in Prompt Changes}
\label{table:inconsistency-types}
\centering
\begin{tabular}{p{0.7\columnwidth}r}
\toprule
\textbf{Category} & \textbf{Count} \\
\midrule
Contradictory Considerations & 8 \\
Structural Conflict & 4 \\
Format Misalignment & 3 \\
\midrule
\textbf{Total} & \textbf{15} \\
\bottomrule
\end{tabular}
\end{table}

\subsection{Theme 4: Prompt Change Impact and Effectiveness}
\subsection*{\textbf{RQ8:} How do prompt changes affect logical consistency in the prompt?}

To understand how prompt changes might introduce logical inconsistencies, we employed a two-stage analysis process. First, we used an LLM with chain-of-thought prompting to identify potentially inconsistent changes, which yielded 205 cases. Two authors then reviewed these cases, confirming 15 instances where prompt changes introduced inconsistencies. See details in section \ref{sec:method}.
% , as summarized in \textit{Table \ref{table:inconsistency-types}.}  
% \hl{See Section X FIX AI:didn't understand this comment.} 

Table \ref{table:inconsistency-types} presents the three categories of inconsistencies we identified. 
% These inconsistencies fall into three distinct categories. 
The first category, \textbf{Contradictory Considerations}, comprises 8 cases where multiple considerations within the same prompt were found to be in direct conflict with each other. For instance, in one prompt, developers included the consideration, ``Chapters should be a meaningful length and not short,'' while later in the prompt adding, ``Keep it short!''. Such opposing considerations make it difficult for the LLM to determine whether the response should prioritize brevity or detail, leading to confusion and reducing the overall effectiveness of the prompt \cite{li2023contradoc}.

The second type of inconsistency, \textbf{Structural Conflicts}, was found in 4 cases. These inconsistencies arise when changes to prompts disrupt their foundational structure, such as through incorrect tag placements or improper formatting. For example, prompts containing instruction tags such as `[INST] Start analysis... [/INST] {input} [/INST]` had closing tags that were misplaced, leading to a broken structure that may confuse the LLM. Such issues prevent the model from correctly interpreting the intended sequence of instructions, resulting in unintended responses.

The third category, \textbf{Format Misalignments}, accounted for 3 cases. These inconsistencies were caused by mismatches between the specified actions in the prompt and their implementation. One instance involved a prompt originally instructing the LLM to ``Buy some movies'' which was later updated to ``Watch some movies'' while retaining the original action tag `[BUY]::`. This inconsistency between the tag and the updated instruction may lead to confusion about whether the model should execute a buying or watching action, potentially causing it to generate unsuitable output.

% \hl{No reference to table III AI: fixed in RQ7's answer}

These findings have important implications for prompt engineering practices. First, they highlight the need for systematic validation approaches during prompt modification, as changes can create ripple effects across different prompt components. Second, they suggest an opportunity for developing automated tools specifically designed to detect logical inconsistencies during prompt development. Third, they emphasize the importance of maintaining version control and documentation for prompts, similar to traditional software artifacts.

\begin{mdframed}[roundcorner=10pt]
\textbf{Observation 8}: Prompt changes can introduce various types of logical inconsistencies, particularly in instruction alignment and structural coherence.%, highlighting the need for systematic validation approaches in prompt engineering.
\end{mdframed}

\subsection*{\textbf{RQ9:} Does the intended prompt change align with the actual outcome when executed by the LLM?}

% Among the seven projects which we ran to analyze the impact of prompt changes on LLM response, in two cases the LLM response were the same for pre- and post-change states. In four projects, the prompt changes ,that were Consideration Modification, were reflected in LLMs' response. In two other projects the response changed as a result of a Consideration Modification change in the prompt. Lastly, in one project, the LLM's response deviated from the expected changes. More specifically, the LLM added a section in its response that were not requested through the changes in the prompt.

To understand how prompt changes affect LLM behavior in practice, we analyzed the alignment between intended changes and actual outcomes. From 101 projects with documented prompt changes, we identified 7 projects with sufficient test cases and example inputs to enable this analysis. While this sample is limited, it provides valuable insights into the practical challenges of prompt modification and testing.

Our analysis revealed three distinct patterns in how LLM responses adapted to prompt changes. 
% \hl{The first pattern entailed .....FIX and do the same for the other patterns; the patterns are buried in the text}
The first pattern entailed accurate adaptation to modifications. In four cases, the responses accurately reflected the modifications, particularly when changes involved adding specific task-related instructions or simplifying existing ones. For example, when a prompt was modified to be ``more straightforward'' and ``focused on the main task,'' the LLM's response appropriately adjusted to these new constraints. The second pattern showed no change in response. In two cases, despite significant prompt modifications such as ``adding few shot examples'', the LLM responses remained unchanged, suggesting that some types of prompt changes may not effectively influence model behavior. The third pattern involved unintended changes. In one case, the response changed but deviated from the intended modification, introducing content not prompted by the changes.

This analysis highlights several critical challenges in prompt engineering for LLM-integrated systems. %Only 7 out of 101 projects had suitable test cases, indicating a significant gap in current testing practices. The presence of runtime-dependent variables and external dependencies further complicates the reconstruction and testing of prompts, emphasizing the need for specialized testing frameworks. Moreover, 
The inconsistent relationship between prompt changes and LLM responses suggests that developers cannot reliably predict how modifications will impact model behavior, underscoring the need for effective validation approaches. Unlike traditional software, prompts require testing methods that accommodate their natural language form and the non-deterministic nature of LLMs \cite{ouyang2024empirical}. Practitioners need systematic review protocols, while researchers have the opportunity to develop automated validation tools that can ensure functional correctness and identify issues in prompt behavior.

% This analysis revealed several critical challenges in prompt engineering. First, the limited availability of test cases (only 7 out of 101 projects had suitable tests) highlights a significant gap in current testing practices for LLM-integrated systems. Second, the presence of runtime-dependent variables and external dependencies made it difficult to reconstruct and test prompts effectively, suggesting a need for better testing frameworks. Third, the inconsistent relationship between prompt changes and LLM responses indicates that developers cannot always predict how modifications will affect model behavior.

% \begin{mdframed}[roundcorner=10pt]
% \textbf{Observation 11}: Prompt modifications do not consistently achieve their intended effects on LLM responses, and the current lack of testing infrastructure makes it difficult to validate changes effectively, highlighting the need for specialized testing frameworks for LLM-integrated systems.
% \end{mdframed}

\begin{mdframed}[roundcorner=10pt]
\textbf{Observation 9}: Prompt modifications do not consistently achieve their intended effects on LLM responses.
\end{mdframed}

% In our analysis of prompt modifications on LLM responses across seven projects, we identified three main patterns: \textbf{1) Unchanged Responses:} In two cases, the LLM responses remained the same before and after the prompt modifications. \textbf{2) Accurate Reflection of Modifications:} In four cases, categorized as Consideration Modifications, the LLM responses accurately incorporated the prompt changes. \textbf{3) Unexpected Behavior:} In one case, the LLM response diverged unexpectedly, adding a section not prompted by the modifications.
% \input{05-discussion}
\section{Threats to Validity}
\label{sec:ttv}

While we employed rigorous methodological approaches in our study, we acknowledge certain considerations regarding validity.

\textbf{External Validity:} %OLD:Our analysis of prompt evolution patterns focused on Python repositories, excluding other popular programming languages, which might exhibit different characteristics. Additionally, our dataset consists of open-source projects on GitHub, which may exhibit different patterns from private or commercial software development. However, our repository selection criteria helps mitigate potential threats by ensuring we analyze established, actively maintained projects with real-world usage of LLM-integrated applications.
Our analysis focused on Python repositories, excluding other languages that may have different characteristics. Additionally, our dataset consists of open-source GitHub projects, which may differ from private or commercial development. However, our selection criteria ensure that we analyze well-established, actively maintained projects with real-world LLM integration, minimizing potential biases.

\textbf{Internal Validity:} %OLD:The main considerations for internal validity stems from our use of automated tools and classification methods. While we used LLMs for commit classification and inconsistency detection, we mitigated this consideration through a comprehensive manual review of the results. Our 15-word minimum threshold for prompt analysis might have excluded some valid but shorter prompts, though this decision was made to ensure meaningful analysis.
The main internal validity considerations arise from using automated tools and classification methods. To address this, we conducted a thorough manual review of results from LLMs used for commit classification and inconsistency detection. Additionally, while our 15-word minimum threshold for prompt analysis may have excluded some shorter but valid prompts, it was set to ensure meaningful analysis.

% For our qualitative analysis process, we addressed potential subjectivity through several measures: employing two independent researchers for coding, conducting multiple rounds of refinement, and calculating inter-rater reliability using Jaccard Index Averaging, which achieved a 65\% agreement rate despite the complexity of multi-label coding.

\textbf{Construct Validity:} Our analysis of prompt effectiveness included 7 out of 101 projects due to the limited availability of suitable test cases, which may not fully represent the relationship between prompt changes and their effects on LLM behavior. Additionally, our definition of inconsistency in prompts might not capture all possible types of conflicts. %Future research could address these limitations by developing standardized testing frameworks for LLM-integrated applications and establishing more comprehensive criteria for identifying prompt inconsistencies.

\section{Conclusion}
\label{sec:conclusion}

In this paper, we presented the first empirical study analyzing prompt evolution in LLM-integrated applications, examining 1,262 prompt changes across 243 GitHub repositories. Our analysis revealed that prompt evolution primarily follows an expansive pattern, with developers favoring addition and modifications over removals, particularly during feature development. We identified critical gaps in current development practices: insufficient documentation, limited testing infrastructure, and the introduction of logical inconsistencies in prompt modifications. These findings highlight the need for specialized tools and practices in prompt engineering, including systematic validation approaches, testing frameworks that account for LLMs' non-deterministic nature, and better documentation practices. As LLM-integrated applications become more prevalent, addressing these challenges will be crucial for ensuring their reliability and maintainability.

\bibliographystyle{IEEEtranS}
\bibliography{icse}

% Generated by IEEEtranS.bst, version: 1.14 (2015/08/26)
\begin{thebibliography}{10}
\providecommand{\url}[1]{#1}
\csname url@samestyle\endcsname
\providecommand{\newblock}{\relax}
\providecommand{\bibinfo}[2]{#2}
\providecommand{\BIBentrySTDinterwordspacing}{\spaceskip=0pt\relax}
\providecommand{\BIBentryALTinterwordstretchfactor}{4}
\providecommand{\BIBentryALTinterwordspacing}{\spaceskip=\fontdimen2\font plus
\BIBentryALTinterwordstretchfactor\fontdimen3\font minus \fontdimen4\font\relax}
\providecommand{\BIBforeignlanguage}[2]{{%
\expandafter\ifx\csname l@#1\endcsname\relax
\typeout{** WARNING: IEEEtranS.bst: No hyphenation pattern has been}%
\typeout{** loaded for the language `#1'. Using the pattern for}%
\typeout{** the default language instead.}%
\else
\language=\csname l@#1\endcsname
\fi
#2}}
\providecommand{\BIBdecl}{\relax}
\BIBdecl

\bibitem{abridge}
\BIBentryALTinterwordspacing
{Abridge AI, Inc.}, ``Abridge: Medical conversation summarization,'' 2021, accessed: 2023-11-03. [Online]. Available: \url{https://www.abridge.com/}
\BIBentrySTDinterwordspacing

\bibitem{10.1145/3674805.3690755}
\BIBentryALTinterwordspacing
L.~Aguiar, M.~Paixao, R.~Carmo, E.~Soares, A.~Leal, M.~Freitas, and E.~Gama, ``Multi-language software development in the llm era: Insights from practitioners’ conversations with chatgpt,'' in \emph{Proceedings of the 18th ACM/IEEE International Symposium on Empirical Software Engineering and Measurement}, ser. ESEM '24.\hskip 1em plus 0.5em minus 0.4em\relax New York, NY, USA: Association for Computing Machinery, 2024, p. 489–495. [Online]. Available: \url{https://doi.org/10.1145/3674805.3690755}
\BIBentrySTDinterwordspacing

\bibitem{semantic_scholar}
\BIBentryALTinterwordspacing
{Allen Institute for AI}, ``Semantic scholar,'' 2015, accessed: 2024-11-03. [Online]. Available: \url{https://www.semanticscholar.org/}
\BIBentrySTDinterwordspacing

\bibitem{Baek2024}
\BIBentryALTinterwordspacing
D.~H. Bæk, ``{GPT Store Statistics \& Facts},'' 2024, accessed: 2024-10-31. [Online]. Available: \url{https://seo.ai/blog/gpt-store-statistics-facts}
\BIBentrySTDinterwordspacing

\bibitem{cai2024demystifying}
Y.~Cai, P.~Liang, Y.~Wang, Z.~Li, and M.~Shahin, ``Demystifying issues, causes and solutions in llm open-source projects,'' \emph{arXiv preprint arXiv:2409.16559}, 2024.

\bibitem{corbin2015basics}
J.~Corbin and A.~Strauss, \emph{Basics of qualitative research}.\hskip 1em plus 0.5em minus 0.4em\relax sage, 2015, vol.~14.

\bibitem{desmond2024exploring}
M.~Desmond and M.~Brachman, ``Exploring prompt engineering practices in the enterprise,'' \emph{arXiv preprint arXiv:2403.08950}, 2024.

\bibitem{dong2022fira}
J.~Dong, Y.~Lou, Q.~Zhu, Z.~Sun, Z.~Li, W.~Zhang, and D.~Hao, ``Fira: fine-grained graph-based code change representation for automated commit message generation,'' in \emph{Proceedings of the 44th International Conference on Software Engineering}, 2022, pp. 970--981.

\bibitem{dubey2024llama}
A.~Dubey, A.~Jauhri, A.~Pandey, A.~Kadian, A.~Al-Dahle, A.~Letman, A.~Mathur, A.~Schelten, A.~Yang, A.~Fan \emph{et~al.}, ``The llama 3 herd of models,'' \emph{arXiv preprint arXiv:2407.21783}, 2024.

\bibitem{feng2010comparison}
L.~Feng, M.~Jansche, M.~Huenerfauth, and N.~Elhadad, ``A comparison of features for automatic readability assessment,'' in \emph{Coling 2010: Posters}, 2010, pp. 276--284.

\bibitem{fernando2023promptbreeder}
C.~Fernando, D.~Banarse, H.~Michalewski, S.~Osindero, and T.~Rockt{\"a}schel, ``Promptbreeder: Self-referential self-improvement via prompt evolution,'' \emph{arXiv preprint arXiv:2309.16797}, 2023.

\bibitem{gao2023makes}
S.~Gao, X.-C. Wen, C.~Gao, W.~Wang, H.~Zhang, and M.~R. Lyu, ``What makes good in-context demonstrations for code intelligence tasks with llms?'' in \emph{2023 38th IEEE/ACM International Conference on Automated Software Engineering (ASE)}.\hskip 1em plus 0.5em minus 0.4em\relax IEEE, 2023, pp. 761--773.

\bibitem{github_copilot_2021}
{GitHub, Inc.}, ``{GitHub Copilot},'' \url{https://github.com/features/copilot}, 2021, accessed: [your access date, e.g., 2024-11-03].

\bibitem{10.1145/3663529.3663849}
\BIBentryALTinterwordspacing
A.~E. Hassan, D.~Lin, G.~K. Rajbahadur, K.~Gallaba, F.~R. Cogo, B.~Chen, H.~Zhang, K.~Thangarajah, G.~Oliva, J.~J. Lin, W.~M. Abdullah, and Z.~M.~J. Jiang, ``Rethinking software engineering in the era of foundation models: A curated catalogue of challenges in the development of trustworthy fmware,'' in \emph{Companion Proceedings of the 32nd ACM International Conference on the Foundations of Software Engineering}, ser. FSE 2024.\hskip 1em plus 0.5em minus 0.4em\relax New York, NY, USA: Association for Computing Machinery, 2024, p. 294–305. [Online]. Available: \url{https://doi.org/10.1145/3663529.3663849}
\BIBentrySTDinterwordspacing

\bibitem{hu2024quantifying}
T.~Hu and N.~Collier, ``Quantifying the persona effect in llm simulations,'' \emph{arXiv preprint arXiv:2402.10811}, 2024.

\bibitem{ibm_watson_assistant}
IBM, ``Ibm watson assistant,'' \url{https://www.ibm.com/products/watsonx-assistant}, accessed: 2024-11-03.

\bibitem{imani2024context}
A.~Imani, I.~Ahmed, and M.~Moshirpour, ``Context conquers parameters: Outperforming proprietary llm in commit message generation,'' \emph{arXiv preprint arXiv:2408.02502}, 2024.

\bibitem{cursor_ai}
\BIBentryALTinterwordspacing
A.~Inc., ``Cursor,'' 2024, version 1.0. [Online]. Available: \url{https://www.cursor.com/}
\BIBentrySTDinterwordspacing

\bibitem{elicit}
\BIBentryALTinterwordspacing
O.~Inc., ``Elicit: The ai research assistant,'' 2021, accessed: 2024-11-03. [Online]. Available: \url{https://elicit.org/}
\BIBentrySTDinterwordspacing

\bibitem{research_rabbit}
\BIBentryALTinterwordspacing
R.~R. Inc., ``Research rabbit,'' 2021, accessed: 2024-11-03. [Online]. Available: \url{https://www.researchrabbit.ai/}
\BIBentrySTDinterwordspacing

\bibitem{tabnine}
\BIBentryALTinterwordspacing
T.~Inc., ``Tabnine: Ai code assistant,'' 2024, accessed: 2024-11-03. [Online]. Available: \url{https://www.tabnine.com/}
\BIBentrySTDinterwordspacing

\bibitem{infermedica}
\BIBentryALTinterwordspacing
{Infermedica}, ``Infermedica: Medical triage and symptom checker,'' 2021, accessed: 2023-11-03. [Online]. Available: \url{https://www.infermedica.com/}
\BIBentrySTDinterwordspacing

\bibitem{jiang2023identifying}
F.~Jiang, Z.~Xu, L.~Niu, B.~Wang, J.~Jia, B.~Li, and R.~Poovendran, ``Identifying and mitigating vulnerabilities in llm-integrated applications,'' \emph{arXiv preprint arXiv:2311.16153}, 2023.

\bibitem{10.1145/2597073.2597074}
\BIBentryALTinterwordspacing
E.~Kalliamvakou, G.~Gousios, K.~Blincoe, L.~Singer, D.~M. German, and D.~Damian, ``The promises and perils of mining github,'' in \emph{Proceedings of the 11th Working Conference on Mining Software Repositories}, ser. MSR 2014.\hskip 1em plus 0.5em minus 0.4em\relax New York, NY, USA: Association for Computing Machinery, 2014, p. 92–101. [Online]. Available: \url{https://doi.org/10.1145/2597073.2597074}
\BIBentrySTDinterwordspacing

\bibitem{kasisto_kai}
Kasisto, ``Kai-gpt: Banking's first purpose-built language model,'' \url{https://kasisto.com/products/kai-gpt/}, accessed: 2024-11-03.

\bibitem{labelstudio}
{Label Studio}, ``{Label Studio},'' \url{https://labelstud.io/}, accessed: 2024-11-03.

\bibitem{landis1977measurement}
J.~R. Landis and G.~G. Koch, ``The measurement of observer agreement for categorical data,'' \emph{biometrics}, pp. 159--174, 1977.

\bibitem{li2024only}
J.~Li, D.~Farag{\'o}, C.~Petrov, and I.~Ahmed, ``Only diff is not enough: Generating commit messages leveraging reasoning and action of large language model,'' \emph{Proceedings of the ACM on Software Engineering}, vol.~1, no. FSE, pp. 745--766, 2024.

\bibitem{li2023contradoc}
J.~Li, V.~Raheja, and D.~Kumar, ``Contradoc: Understanding self-contradictions in documents with large language models,'' \emph{arXiv preprint arXiv:2311.09182}, 2023.

\bibitem{li2023spell}
Y.~B. Li and K.~Wu, ``Spell: Semantic prompt evolution based on a llm,'' \emph{arXiv preprint arXiv:2310.01260}, 2023.

\bibitem{liang2024prompts}
J.~T. Liang, M.~Lin, N.~Rao, and B.~A. Myers, ``Prompts are programs too! understanding how developers build software containing prompts,'' \emph{arXiv preprint arXiv:2409.12447}, 2024.

\bibitem{10.1145/3597503.3608128}
\BIBentryALTinterwordspacing
J.~T. Liang, C.~Yang, and B.~A. Myers, ``A large-scale survey on the usability of ai programming assistants: Successes and challenges,'' in \emph{Proceedings of the IEEE/ACM 46th International Conference on Software Engineering}, ser. ICSE '24.\hskip 1em plus 0.5em minus 0.4em\relax New York, NY, USA: Association for Computing Machinery, 2024. [Online]. Available: \url{https://doi.org/10.1145/3597503.3608128}
\BIBentrySTDinterwordspacing

\bibitem{lin2004rouge}
C.-Y. Lin, ``Rouge: A package for automatic evaluation of summaries,'' in \emph{Text summarization branches out}, 2004, pp. 74--81.

\bibitem{liu2020atom}
S.~Liu, C.~Gao, S.~Chen, L.~Y. Nie, and Y.~Liu, ``Atom: Commit message generation based on abstract syntax tree and hybrid ranking,'' \emph{IEEE Transactions on Software Engineering}, vol.~48, no.~5, pp. 1800--1817, 2020.

\bibitem{min2022rethinking}
S.~Min, X.~Lyu, A.~Holtzman, M.~Artetxe, M.~Lewis, H.~Hajishirzi, and L.~Zettlemoyer, ``Rethinking the role of demonstrations: What makes in-context learning work?'' \emph{arXiv preprint arXiv:2202.12837}, 2022.

\bibitem{nahar2024beyond}
N.~Nahar, C.~K{\"a}stner, J.~Butler, C.~Parnin, T.~Zimmermann, and C.~Bird, ``Beyond the comfort zone: Emerging solutions to overcome challenges in integrating llms into software products,'' \emph{arXiv preprint arXiv:2410.12071}, 2024.

\bibitem{nie2021coregen}
L.~Y. Nie, C.~Gao, Z.~Zhong, W.~Lam, Y.~Liu, and Z.~Xu, ``Coregen: Contextualized code representation learning for commit message generation,'' \emph{Neurocomputing}, vol. 459, pp. 97--107, 2021.

\bibitem{nuance_dax}
\BIBentryALTinterwordspacing
{Nuance Communications, Inc.}, ``Nuance dax: Ambient clinical intelligence,'' 2021, accessed: 2023-11-03. [Online]. Available: \url{https://www.nuance.com/healthcare/ambient-clinical-intelligence.html}
\BIBentrySTDinterwordspacing

\bibitem{ouyang2024empirical}
S.~Ouyang, J.~M. Zhang, M.~Harman, and M.~Wang, ``An empirical study of the non-determinism of chatgpt in code generation,'' \emph{ACM Transactions on Software Engineering and Methodology}, 2024.

\bibitem{pickerill2020phantom}
P.~Pickerill, H.~J. Jungen, M.~Ochodek, M.~Ma{\'c}kowiak, and M.~Staron, ``Phantom: Curating github for engineered software projects using time-series clustering,'' \emph{Empirical Software Engineering}, vol.~25, pp. 2897--2929, 2020.

\bibitem{pister2024promptset}
K.~Pister, D.~J. Paul, I.~Joshi, and P.~Brophy, ``Promptset: A programmer's prompting dataset,'' in \emph{Proceedings of the 1st International Workshop on Large Language Models for Code}, 2024, pp. 62--69.

\bibitem{real1996probabilistic}
R.~Real and J.~M. Vargas, ``The probabilistic basis of jaccard's index of similarity,'' \emph{Systematic biology}, vol.~45, no.~3, pp. 380--385, 1996.

\bibitem{sahoo2024systematic}
P.~Sahoo, A.~K. Singh, S.~Saha, V.~Jain, S.~Mondal, and A.~Chadha, ``A systematic survey of prompt engineering in large language models: Techniques and applications,'' \emph{arXiv preprint arXiv:2402.07927}, 2024.

\bibitem{sauvola2024future}
J.~Sauvola, S.~Tarkoma, M.~Klemettinen, J.~Riekki, and D.~Doermann, ``Future of software development with generative ai,'' \emph{Automated Software Engineering}, vol.~31, no.~1, p.~26, 2024.

\bibitem{schulhoff2024prompt}
S.~Schulhoff, M.~Ilie, N.~Balepur, K.~Kahadze, A.~Liu, C.~Si, Y.~Li, A.~Gupta, H.~Han, S.~Schulhoff \emph{et~al.}, ``The prompt report: A systematic survey of prompting techniques,'' \emph{arXiv preprint arXiv:2406.06608}, 2024.

\bibitem{shin2023prompt}
J.~Shin, C.~Tang, T.~Mohati, M.~Nayebi, S.~Wang, and H.~Hemmati, ``Prompt engineering or fine tuning: An empirical assessment of large language models in automated software engineering tasks,'' \emph{arXiv preprint arXiv:2310.10508}, 2023.

\bibitem{strauss1998basics}
A.~Strauss and J.~Corbin, ``Basics of qualitative research techniques,'' 1998.

\bibitem{10.1145/2393596.2393656}
\BIBentryALTinterwordspacing
Y.~Tao, Y.~Dang, T.~Xie, D.~Zhang, and S.~Kim, ``How do software engineers understand code changes? an exploratory study in industry,'' in \emph{Proceedings of the ACM SIGSOFT 20th International Symposium on the Foundations of Software Engineering}, ser. FSE '12.\hskip 1em plus 0.5em minus 0.4em\relax New York, NY, USA: Association for Computing Machinery, 2012. [Online]. Available: \url{https://doi.org/10.1145/2393596.2393656}
\BIBentrySTDinterwordspacing

\bibitem{tryggeseth1997report}
E.~Tryggeseth, ``Report from an experiment: Impact of documentation on maintenance,'' \emph{Empirical Software Engineering}, vol.~2, no.~2, pp. 201--207, 1997.

\bibitem{10555619}
R.~Tufano, A.~Mastropaolo, F.~Pepe, O.~Dabić, M.~Di~Penta, and G.~Bavota, ``Unveiling chatgpt’s usage in open source projects: A mining-based study,'' in \emph{2024 IEEE/ACM 21st International Conference on Mining Software Repositories (MSR)}, 2024, pp. 571--583.

\bibitem{Weber2024LargeLM}
\BIBentryALTinterwordspacing
I.~Weber, ``Large language models as software components: A taxonomy for llm-integrated applications,'' \emph{ArXiv}, vol. abs/2406.10300, 2024. [Online]. Available: \url{https://api.semanticscholar.org/CorpusID:270559976}
\BIBentrySTDinterwordspacing

\bibitem{wei2022chain}
J.~Wei, X.~Wang, D.~Schuurmans, M.~Bosma, F.~Xia, E.~Chi, Q.~V. Le, D.~Zhou \emph{et~al.}, ``Chain-of-thought prompting elicits reasoning in large language models,'' \emph{Advances in neural information processing systems}, vol.~35, pp. 24\,824--24\,837, 2022.

\bibitem{White2023APP}
\BIBentryALTinterwordspacing
J.~White, Q.~Fu, S.~Hays, M.~Sandborn, C.~Olea, H.~Gilbert, A.~Elnashar, J.~Spencer-Smith, and D.~C. Schmidt, ``A prompt pattern catalog to enhance prompt engineering with chatgpt,'' \emph{ArXiv}, vol. abs/2302.11382, 2023. [Online]. Available: \url{https://api.semanticscholar.org/CorpusID:257079092}
\BIBentrySTDinterwordspacing

\bibitem{wu2023bloomberggpt}
S.~Wu, O.~Irsoy, S.~Lu, V.~Dabravolski, M.~Dredze, S.~Gehrmann, P.~Kambadur, D.~Rosenberg, and G.~Mann, ``Bloomberggpt: A large language model for finance,'' \emph{arXiv preprint arXiv:2303.17564}, 2023.

\bibitem{yang2024sifid}
J.~Yang, H.~Liu, W.~Guo, Z.~Rao, Y.~Xu, and D.~Niu, ``Sifid: Reassess summary factual inconsistency detection with llm,'' \emph{arXiv preprint arXiv:2403.07557}, 2024.

\bibitem{zhang2024detecting}
Y.~Zhang, ``Detecting code comment inconsistencies using llm and program analysis,'' in \emph{Companion Proceedings of the 32nd ACM International Conference on the Foundations of Software Engineering}, 2024, pp. 683--685.

\end{thebibliography}

\end{document}